\documentclass[conference]{IEEEtran}
\usepackage{ifpdf}
\usepackage{cite}
\usepackage{graphicx}

\usepackage{amsmath, amssymb}
\usepackage{algorithm, algorithmic}
\usepackage{array}

\usepackage{stfloats}
\usepackage{url}
\usepackage{multirow}
\usepackage[table,xcdraw]{xcolor}
\usepackage{longtable}
\usepackage{threeparttable}
\usepackage{lscape}
\usepackage{verbatim}
\usepackage{bbding} %×

\usepackage{marvosym}
\usepackage{ulem}

\usepackage[colorlinks,
linkcolor=red,
anchorcolor=green,
citecolor=blue]{hyperref}

\usepackage{fancyhdr}
\newcommand{\Rone}[1]{\textcolor{black}{#1}}

\newcommand{\Review}[1]{\textcolor{black}{#1}}
\newcommand{\fix}[1]{\textcolor{black}{#1}}
\newcommand{\update}[1]{\textcolor{black}{#1}}
\newcommand{\refine}[1]{\textcolor{black}{#1}}
\newcommand{\rOneFC}[1]{\textcolor{black}{#1}}
\newcommand{\modify}[1]{\textcolor{black}{#1}}
\newcommand{\final}[1]{\textcolor{black}{#1}}
\newcommand{\finetune}[1]{\textcolor{black}{#1}}
\newcommand{\publish}[1]{\textcolor{black}{#1}}
\newcommand\Mark[1]{\textsuperscript#1}

\begin{document}

% 添加版权声明
% \IEEEoverridecommandlockouts


% \title{BTA:  Energy-Efficient and Precision-Scalable Accelerator for Binary Transformer-based Neural Networks}
% \title{An Energy-Efficient and Precision-Scalable Accelerator for Binary Transformer-based Neural Networks}
% \title{An Edge-Friendly and Accuracy-Scalable Accelerator for Binary Transformer-based Neural Networks}
% \title{EBTA: An Edge-Friendly Accelerator for \\ Binary Transformer-based Neural Networks}
% \title{BETA: Binary Edge Transformer Accelerator on FPGA}
\title{BETA: \uline{B}inarized \uline{E}nergy-Efficient \uline{T}ransformer \uline{A}ccelerator at the Edge}
% \title{BETA: Enabling Binary Edge Transformer Acceleration on FPGA}
% [Note] Chao Fang: 是否考虑给工作取个缩写名？
% [Note] BTA: Binary Transformer Accelerator
% [Note] Title太长，两行放不下了哇

\author{
    Yuhao~Ji\Mark{1}, Chao~Fang\Mark{1}, and Zhongfeng~Wang\Mark{1}$^{,}$\Mark{2}$^{(\textrm{\Letter})}$ \\
	\IEEEauthorblockA{
		\Mark{1}School of Electronic Science and Engineering, 
		Nanjing University, Nanjing, China\\
            \Mark{2}School of Integrated Circuits, Sun Yat-sen University, Shenzhen, China\\
		Email:
		\{201180131, fantasysee\}@smail.nju.edu.cn, zfwang@nju.edu.cn
    }
	% \thanks{Zhongfeng Wang is the corresponding author. This work was supported in part by the National Natural Science Foundation of China under Grant 62174084, 62104097, in part by the High-Level Personnel Project of Jiangsu Province under Grant JSSCBS20210034, and in part by Postgraduate Research \& Practice Innovation Program of Jiangsu Province under Grant No. 149.
	% }
}% <-this % stops a space

\maketitle
% 添加版权声明
% \IEEEpubidadjcol
% \input{0-abstr.tex}

\begin{abstract}
% \Rone{binary Transformers are promising in edge deployment application due to its small parameter size and high computation efficiency.}
% \Rone{Existing works have achieved binarization for transformer-based model parameters and activations while maintaining a considerable accuracy.}
% \fix{Existing binary Transformers are promising in edge deployment due to their small parameter size, low computational complexity and considerable accuracy.}
\fix{Existing binary Transformers are promising in edge deployment due to their \rOneFC{compact model size}, low computational complexity, and \rOneFC{considerable inference accuracy}.}
% 和Intro para2对应
% \Rone{However, hardware implementation of binary Transformers requires further exploration, especially for the quantized matrix multiplication (QMM), which is the key operation in Transformer inference.}
% \fix{However, the implementation efficiency and its balance with accuracy of binary Transformers are not fully explored by traditional processors or accelerators.}
% \update{However, the implementation efficiency and its balance with model accuracy are not fully explored by traditional processors or accelerators.}
\rOneFC{However, deploying binary Transformers faces challenges on prior processors due to inefficient execution of quantized matrix multiplication (QMM) and the energy consumption overhead caused by multi-precision activations.}
% \Rone{To address this issue, we first propose a computation flow abstraction method for binary Transformer-based Neural Networks to reduce the computational complexity, and then corresponding accelerator, named BETA, is designed targeting on resource-limited edge devices.}
% \fix{To address this issue, we first propose a general computation flow abstraction method, adjusting the calculation order to reduce full-precision operations, and then corresponding accelerator for abstracted flow, named BETA, is designed targeting on resource-constrained edge devices.}
% \update{To address this issue, we first propose a general computation flow abstraction method, adjusting the calculation order to reduce full-precision operations.}
\rOneFC{To tackle the challenges above, we first develop a computation flow abstraction method for binary Transformers to improve QMM execution efficiency by optimizing the computation order.}
\rOneFC{Furthermore, a binarized energy-efficient Transformer accelerator, namely BETA, is proposed to boost the efficient deployment at the edge.}
\rOneFC{Notably, BETA features a configurable QMM engine, accommodating diverse activation precisions of binary Transformers and offering high-parallelism and high-speed \modify{for QMMs} with impressive energy efficiency.}
\rOneFC{Experimental results evaluated on ZCU102 FPGA show BETA achieves an average energy efficiency of 174 GOPS/W, which is \modify{1.76}$\sim$\modify{21.92}$\times$ higher than prior FPGA-based accelerators, showing BETA's good potential for edge Transformer acceleration.}
% \TODO{fill the experimental data, Gops/DSP maybe}
% \Rone{Moreover, BETA has the ability to dynamically adjust the hardware efficiency and model accuracy.}
% \Rone{All advantages above make BETA an excellent architecture suitable for edge deployment.}
% \fix{Based on the advantages above, we believe BETA is a competent choice for the edge deployment of Transformer.}
\end{abstract}
% \input{1-intro.tex}
\vspace{-1em}
\section{Introduction} \label{sec:intro}
% Posit \cite{gustafson2017beating} is regarded as a drop-in replacement for traditional floating-point numbers due to its better trade-off between dynamic range and accuracy.
% \Rone{Transformer-based neural networks\cite{Attention} are excellent deep learning models which have been used across a wide range of applications in natural language process (NLP)\cite{BERT}, computer vision (CV)\cite{BEiT}, and speech recognition.}
% \fix{Transformer-based neural networks\cite{Attention} are excellent deep learning models which have been used across a wide range of applications in natural language process (NLP)\cite{BERT} and computer vision (CV)\cite{BEiT}.}
% \Rone{Due to their superior accuracy, many works are dedicated to deploying Transformer-based neural networks on edge devices, such as smart phones and embedded systems, to perform different tasks in real time, such as voice recognition, text translation, smart assistant, etc.}
% \fix{Due to their superior accuracy, many works are dedicated to deploying Transformer-based neural networks on edge devices\cite{ViTA}, such as smart phones and embedded systems, to perform different tasks without transmitting data to the cloud, such as voice recognition, text translation, smart assistant, etc.}
% \refine{Due to their superior accuracy, many works are dedicated to deploying Transformer-based neural networks on edge devices\cite{ViTA}, such as smart phones and embedded systems, to perform different tasks, such as voice recognition, text translation, smart assistant, etc., without transmitting data to the cloud.}
\rOneFC{In recent years, large language models (LLMs) \cite{brown2020language} have seen a surge in popularity, with applications ranging from natural language understanding \cite{DevlinCLT19} and generation \cite{zhou2023solving} to computer vision \cite{chen2022visualgpt} and robotics \cite{singh2023progprompt}.}
\rOneFC{Transformer-based neural networks \cite{vaswani2017attention} have become the backbone of many LLMs.}
% \rOneFC{However, deploying Transformers on resource-constrained edge devices, such as mobile phones and wearables, remains challenging due to their computational and memory demands. Striking a balance between model efficiency and accuracy is essential for these applications at the edge deployment.}
\publish{However, deploying Transformers on resource-constrained edge devices, such as mobile phones and wearables, remains challenging due to their computational and memory demands.}
% \Rone{Edge deployment eliminates the need to transmit data to the cloud, resulting in reduced communication time, improved response speed, lower latency, and enhanced privacy and security.}

% \Rone{However, the Transformer network's large parameter size and computational requirements make it challenging to deploy on resource-constrained devices such as smartphones and smartwatches, which have limited computing power and storage capacity.
% }
% \Rone{However, the Transformer networks' large parameter size and computational requirements\cite{FullStack} make it challenging to deploy on resource-constrained edge devices, which have limited storage capacity and computing resources.}
% \refine{However, the Transformer networks' large model size and computational requirements\cite{FullStack} make them challenging to deploy on edge devices, which have limited storage capacity and computing resources.}
% [Note]：直接提量化, 不提其它的模型压缩方法, 节约篇幅
% 怎么把energy相关加进去？
% \Rone{To address this issue, various quantization methods\cite{Q8BERT, TernaryBERT, BEBERT, BinaryBERT, BiT, BiBERT} have been proposed, which partially use lower numerical precision for calculations while maintaining acceptable performance degradation.}
\Rone{To address this issue, various quantization approaches \cite{Q8BERT, TernaryBERT, BEBERT, BinaryBERT, BiT, BiBERT, le2023binaryvit, UCViT} have been proposed, which partially use lower numerical precision for calculations while maintaining \rOneFC{satisfying model accuracy}.}
% \Rone{Notably, when quantized to 1-bit, also known as binarization, parameters can be represented in only 1-bit width, and computations can be reduced to bitwise operations, minimizing both parameter storage and computational complexity.}
\rOneFC{Notably, when model parameters are quantized to 1-bit, also known as binarization, computations can be reduced to bitwise operations, minimizing both parameter storage and computational complexity.}
% \Rone{Compared to full-precision models, binary Transformer networks theoretically offer a 32x compression ratio and 64x multiplication speedup ratio, significantly alleviating the deployment pressure on edge devices.}
% \fix{Compared to FIX-32 full-precision models, binary Transformer networks theoretically offer a 32x compression ratio, significantly alleviating the deployment pressure on edge devices.}
% \update{Compared to 32-bit fixed-point (FIX-32) full-precision models, binary Transformer networks theoretically offer a 32x compression ratio, significantly alleviating the deployment pressure on edge devices.}
% \refine{Compared to 32-bit floating-point (FP-32) or 16-bit fixed-point (FIX-16) full-precision models, binary Transformers theoretically offer a 32x or 16x compression ratio, respectively, significantly alleviating the deployment pressure on edge devices.}
% \refine{Compared to 32-bit floating-point (FP-32) or 16-bit fixed-point (FIX-16) \rOneFC{full-precision} models, binary Transformers theoretically offer a 32x or 16x compression ratio, respectively, alleviating \rOneFC{the computational and storage requirements significantly for deploying models on edge devices.}}
\refine{Compared to 32-bit floating-point (FP-32) or 16-bit fixed-point (FIX-16) \rOneFC{full-precision} models, binary Transformers theoretically offer a 32\modify{$\times$} or 16\modify{$\times$} compression ratio, respectively, alleviating \rOneFC{the computational and storage requirements significantly for deploying models on edge devices.}}
% \refine{Compared to \modify{32-bit floating-point (FP-32) full-precision models}, binary Transformer networks theoretically offer a \modify{32x} compression ratio, significantly alleviating the deployment pressure on edge devices.}
% \Rone{Mainstream binary Transformers like BiT \cite{BiT} have achieved impressive results, with an accuracy loss of only 5.4\% and a model compression rate of 31.2x.}
% \rOneFC{For instance, BiT \cite{BiT} have achieved a model compression rate of 31.2x with a negligible accuracy loss of only 5.4\% for edge deployment.}
\rOneFC{For instance, BiT \cite{BiT} have achieved a model compression ratio of 31.2\modify{$\times$} with a negligible accuracy loss of only 5.4\% for edge deployment.}
% MNLI-m Accuracy: BERT:84.9%, BiT W1A1:79.5% 5.4%
% SIZE: BERT:418MB, BiT W1A1: 13.4MB    31.2x
% \Rone{While quantization methods have achieved significant results in reducing parameters and computations, deploying binary Transformers on existing edge devices still presents challenges.}
% \Rone{However, deploying binary Transformers on existing edge devices still presents challenges.}
\refine{However, edge deployment of binary Transformers still presents challenges.}
% \Rone{First, traditional processors or accelerators are primarily optimized for full-precision operations, and the key calculations required for binary Transformers, such as QMM, cannot be efficiently executed on them.}
% \update{First, traditional processors or accelerators are primarily optimized for full-precision operations, and the key calculations required for binary Transformers, such as quantized matrix multiplication (QMM), cannot be efficiently executed on them.}
% \refine{First, traditional processors or accelerators are mostly optimized for full-precision or moderately quantized models, and the key calculations required for binary Transformers, such as quantized matrix multiplication (QMM), cannot be efficiently executed on them.}
% \refine{First, traditional processors or accelerators are mostly optimized for full-precision or moderately quantized models, and the key calculations required for binary Transformers, the quantized matrix multiplication (QMM), cannot be efficiently executed on them.}
\refine{First, prior processors or accelerators \cite{Lu, ViA, SwiftTron, STA, EFA-Trans, FTRANS, ViTA, GOBO, A3, fang2022efficient, Rodrigue2022Resource-Saving} are mostly optimized for full-precision or moderately quantized models, and the key calculations required for binary Transformers, \modify{two types of} quantized matrix multiplication (QMM), \modify{i.e. activation$\times$weight and activation$\times$activation}, cannot be efficiently executed on them.}
% \Rone{This hinders the binary Transformers from fully utilizing hardware parallelism and vectorization capabilities, restricting its acceleration performance.}
% \Rone{Also, existing binary Transformer designs always sacrifice inference accuracy to achieve higher speed and efficiency. However, in some applications, high model accuracy remains crucial. The current design methods have not found an optimal balance between providing efficient computational acceleration and maintaining acceptable inference accuracy.}
% \update{Second, to maintain high accuracy, many binary Transformer designs have various versions of different activation precisions.}
% \modify{Second, to maintain high accuracy, binary Transformers may have several versions of different activation precisions.}
% \modify{Second, to maintain high accuracy, knowledge distillation technique is commonly applied in binary Transformers training, thus generating several network versions of different activation precisions.}
\modify{Second, to meet different edge scenarios with distinct energy efficiency and accuracy demands, it is necessary to deploy binary Transformers of different activation precisions \cite{BiT, BinaryBERT}.}
% \update{Multi-precision activations multiplication with no binary parameter involved increases the energy consumption overhead of binary Transformer accelerator design.}
% \update{Both the challenges mentioned above are new problems compared to traditional binary Convolutional Neural Networks (CNN) accelerator design.}
% \update{Multi-precision activations multiplication with no binary parameter involved potentially increases the energy consumption overhead of binary Transformer accelerator design, which is a new issue compared to binary Convolutional Neural Networks (CNN) accelerator design.}
% \modify{Multi-precision activations multiplication with no binary parameter involved potentially increases the energy consumption overhead, which is a new issue compared to binary Convolutional Neural Networks (CNN) accelerator design.}
\modify{Multi-precision activations multiplication with no binary parameter involved potentially increases the energy consumption overhead.}

% \Rone{To tackle the above challenges, in this paper, we propose BETA, an Edge-Friendly binary Transformer-based Neural Networks Accelerator capable of performing specially abstract computation flow of binary Transformer inference efficiently.}
% \fix{To tackle the above challenges, in this paper, we propose BETA, an Edge-Friendly binary Transformer-based Neural Networks Accelerator capable of performing binary Transformer inference efficiently.}
% \fix{To tackle the above challenges, in this paper, we first develop a general computation flow abstraction method suitable for all binary Transformers to reduce the number of full-precision operations by adjusting the computation order.}
\fix{To tackle the above challenges, in this paper, we first develop a general computation flow abstraction method \modify{for binary Transformers} to reduce the number of full-precision operations by optimizing the computation order.}
% \fix{Corresponding accelerator named BETA, an Edge-Friendly Binary Transformer-based Neural Networks Accelerator, is proposed, which is capable of performing abstract computation flow of binary Transformer inference efficiently.}
% \fix{Corresponding accelerator named BETA, \modify{Binarized Energy-Efficient Transformer Accelerator at the Edge}, is proposed, \modify{enabling efficient abstract binary Transformer deployment at the edge}.}
\rOneFC{On top of that, we propose a binarized energy-efficient Transformer accelerator, namely BETA, \modify{enabling efficient binary Transformer deployment at the edge.}}
% \Rone{BETA significantly improves the computation throughput while obtaining the scalability by applying unfolding technique to achieve high parallelisim and bit-serial method to support multi-precision operation.}
% \Rone{BETA significantly improves QMM computation throughput and reduces the latency by applying unfolding technique to achieve high parallelism and optimizing the accumulation structure to shorten critical path.}
% \fix{QMM is the computing core in binary Transformers, so we specially design a QMM Engine in BETA with high throughput and low latency by applying unfolding technique to achieve high parallelism and optimizing the accumulation structure to shorten critical path.}
% \fix{QMM is the computing core in binary Transformers, so we specially design \modify{a high-throughput QMM Engine} in BETA, applying unfolding technique to achieve high parallelism and optimizing the accumulation structure to \modify{reduce circuit delay}.}
\rOneFC{To improve the performance of QMM, which are \publish{the} dominated operations in binary Transformers, we design a high-throughput QMM engine in BETA. This engine leverages the unfolding technique to achieve high parallelism and optimizes the accumulation structure to reduce datapath latency.}
% \Rone{Additionally, QMM Engine introduces data-packing, which allows the accelerator to perform different activation precision versions of binary Transformers.}
% \fix{Additionally, we propose a fixed input-width design, based on which data-packing is introduced to improve the hardware utilization, allowing the accelerator to perform different activation precision versions of binary Transformers.}
% \update{Additionally, we propose a fixed input-width and configurable PE design with high hardware utilization, allowing the accelerator to flexibly perform different activation precision versions of binary Transformers as needed.}
% \update{Additionally, we propose a fixed input-width and configurable PE design with high hardware utilization, which is able to flexibly process inputs in different activation precision versions of binary Transformers as needed.}
% \update{Additionally, we propose a fixed input-width and configurable PE design with high hardware utilization, \modify{flexibly processing diverse activation precisions of binary Transformers with impressive energy efficiency}.}
\update{Additionally, we propose a configurable PE design, \modify{flexibly processing diverse activation precisions of binary Transformers with impressive energy efficiency}.}
% \Rone{Therefore, EBAT has great potential to achieve superior performance and scalability on edge devices.} 
% \Rone{According to the given architecture innovations, BETA has great potential as the edge devices Transformer accelerator with superior performance and configurability.} 
% \modify{According to the proposed innovations, BETA has good potential as the edge Transformer accelerator with superior performance and configurability.} 
% \rOneFC{BETA has good potential as the edge Transformer accelerator with superior performance and configurability.}

\Review{To summarize, our contributions are as follows.}
\begin{enumerate}
    \item 
    \modify{We abstract the computation process involved in binary Transformers by optimizing the computation order and fusing full-precision coefficients and offsets, which results in reduced computational complexity and significant energy savings without \rOneFC{impacts on model accuracy}.}
    % \refine{We abstract the computation process involved in binary Transformers. By adjusting the computation order and fusing full-precision coefficients and offsets, we reduce computational complexity and achieve \modify{282.64x energy efficiency improvement over \modify{unabstract} FP-32 designs.}}
    \item 
    \rOneFC{We propose BETA, a novel architecture to efficiently deploy binary Transformers. To the best of our knowledge, BETA is the first dedicated accelerator to support diverse activation precisions of binary Transformers. It achieves an average energy efficiency of 174 GOPS/W on ZCU102 FPGA, which is \modify{1.76}$\sim$\modify{21.92}$\times$ higher than prior FPGA-based accelerators \cite{ViTA, STA, EFA-Trans, VAQF}.}
    \item 
    \rOneFC{We design a high-parallelism, high-speed QMM engine that performs two types of QMM and accommodates various activation precisions, enabling dynamic adjustment between computational efficiency and model accuracy to meet different application demands at the edge.}
\end{enumerate}
\section{Background and Motivation}\label{sec:bkg}
\begin{figure}[ht]
    \centering
    \includegraphics[width=0.44\textwidth]{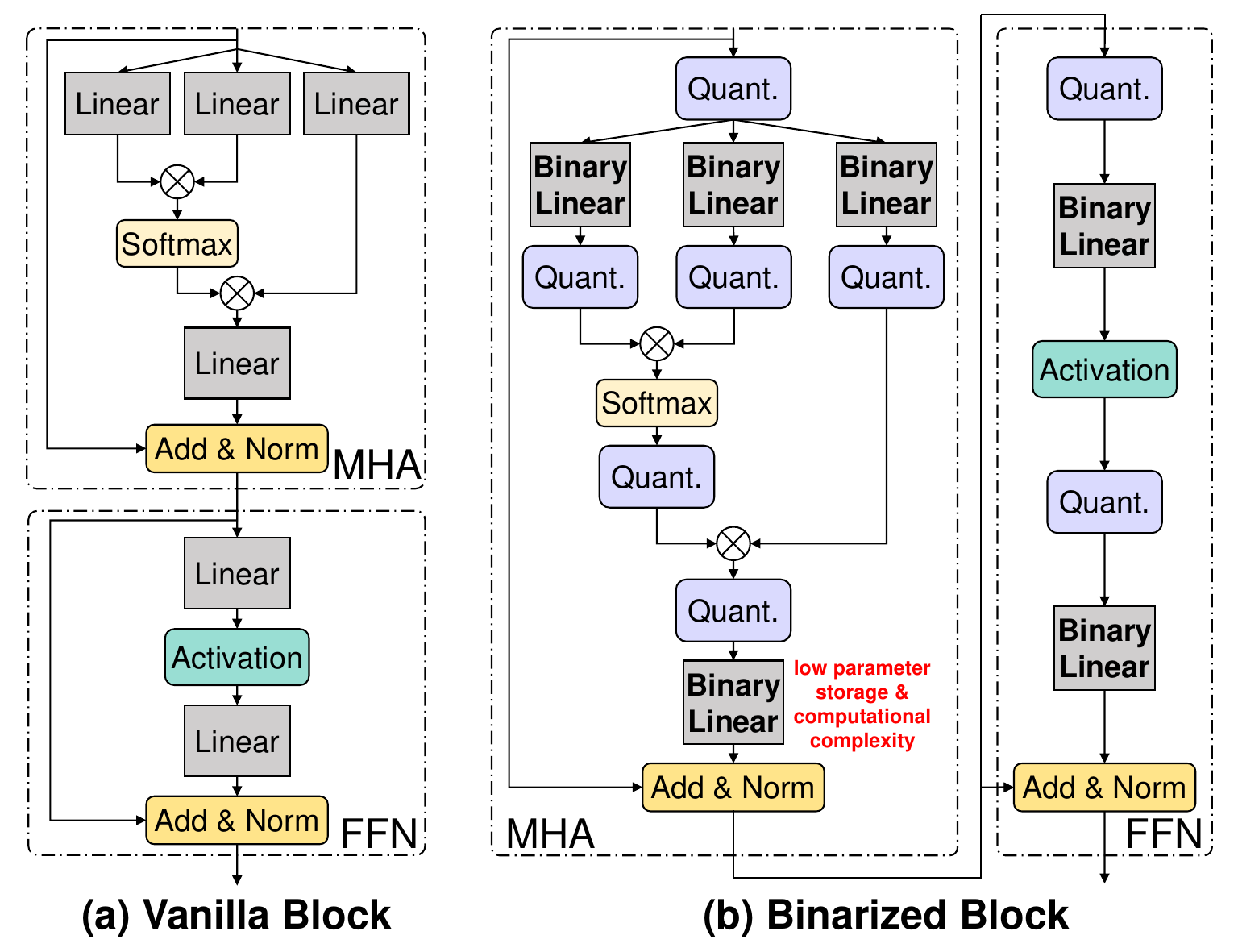}
    % \caption{Overview of traditional Transformer block (left) and Binary Transformer block (right).}
    % \caption{Overview of traditional Transformer block (a) and Binary Transformer block (b).}
    \caption{\rOneFC{Overview of MHA and FFN blocks in (a) vanilla Transformer and (b) binary Transformer, respectively.}}
    \label{Transformer_Arch}
    \vspace{-0.3cm}
\end{figure}

% \Rone{A Transformer network is composed of an embedding layer, several Transformer blocks, and a linear output layer used for classification or regression.}
% \Rone{Each Transformer block includes a multi-head attention (MHA) layer followed by a feed-forward network (FFN), as illustrated in Fig.~\ref{Transformer_Arch}, where the computationally intensive matrix multiplication (MM) occupies a significant portion of the overall computational workload.}
\rOneFC{The main structure of a Transformer is a stack of Transformer blocks, each of which consists of multi-head attention (MHA) blocks and feed-forward network (FFN) blocks.}
\rOneFC{Fig.~\ref{Transformer_Arch} presents the details of MHA and FFN blocks in vanilla and binary Transformers.}
\rOneFC{Compared to the vanilla Transformer, the binary Transformer incorporates binary weights and quantized activations, resulting in low parameter storage and computational complexity.}

\fix{Currently, most Transformer hardware accelerators are optimized for full-precision\cite{Lu, ViA, Rodrigue2022Resource-Saving} or moderately quantized\cite{SwiftTron, STA, EFA-Trans, FTRANS, ViTA, GOBO, A3} Transformers.}
% \modify{ViA\cite{ViA} optimizes hardware architecture based on the characteristics of Vision Transformer (ViT).}
% \modify{For example, ViA\cite{ViA} proposes a novel Vision Transformer (ViT) hardware architecture in FP-16 format based on the characteristics of ViT.}
% \finetune{For instance, ViA\cite{ViA} introduces a novel hardware architecture for Vision Transformer (ViT) acceleration in FP-16 format, aligning with ViT's inherent characteristics.}
\rOneFC{For instance, ViA\cite{ViA} presents a novel hardware architecture tailored for accelerating Vision Transformers (ViT) in the FP-16 format.}
% \modify{STA\cite{STA} presents an algorithm-hardware co-optimized framework to study the efficiency of N:M sparse Transformers in FIX-16 format.}
% \finetune{STA\cite{STA} introduces an algorithm-hardware co-optimized framework, realizing flexible and efficient deployment of FIX-16 Transformers by leveraging general N:M sparsity patterns.}
\rOneFC{STA\cite{STA} develops an algorithm-hardware co-optimized framework that enables flexible and efficient deployment of FIX-16 Transformers by harnessing general N:M sparsity patterns.}
% \modify{EFA-Trans\cite{EFA-Trans} proposes a dense-sparse compatible designs, enabling configurable and efficient implementation of Transformers, while quantizing the networks to FIX-8.}
\finetune{EFA-Trans\cite{EFA-Trans} proposes a hardware design for FIX-8 Transformer models, which is compatible with both dense and sparse configurations.}
% \modify{VAQF\cite{VAQF} fully exploits the speedup potential of binarization by turning multiplication to bit-wise operation.}
\fix{Deploying binary Transformers on these accelerators leads to a huge waste of resources, resulting in low energy efficiency.}
% \final{Notably, VAQF\cite{VAQF} develops a binary ViT accelerator generator, which fully exploits the speedup potential of binarization by turning multiplication to bit-wise operation.}
% \final{However, the generated accelerator only supports one activation precision in each compilation and the second type of QMM, activation$\times$activation, is not considered in the design.}
% \final{However, the generated accelerator merely executes the first type of QMM, i.e. activation$\times$weight, with only one activation precision support in each compilation.}
% \final{However, the generated accelerator merely \finetune{supports} the first type of QMM, i.e. activation$\times$weight, with only one activation precision support in each compilation.}
% \rOneFC{Notably, VAQF\cite{VAQF} presents a binary ViT accelerator generator that fully exploits the speedup potential of binarization by turning multiplication into bit-wise operation. However, the generated accelerator only supports one activation precision in each compilation and does not consider the second type of QMM, activation$\times$activation.}
\rOneFC{Notably, VAQF\cite{VAQF} presents a binary ViT accelerator generator that fully exploits the speedup potential of binarization by turning multiplication into bit-wise operation. However, the generated accelerator only supports one activation precision in each compilation and does not consider the \publish{QMM of activation$\times$activation}.}
\rOneFC{BETA differs from previous works mainly in two aspects: 1) BETA is dedicated \publish{for} binary Transformers, and a general computation flow abstraction method is proposed to further reduce the computational complexity. 2) BETA theoretically supports all binary Transformers, including two types of QMM equipped with various activation precisions, which can be flexibly configured on-the-fly.}
% \fix{To the best of our knowledge, this is the first dedicated binary Transformer accelerator with scalable activation precision support.}

\section{Hardware Acceleration}\label{sec：design}
% In the following, we first discuss the computation flow abstraction of binary Transformers to facilitate the hardware implementation. Then the overall architecture of BETA is shown. Moreover, we present the details of QMM engine component to further elaborate the advantages of our design. 
\subsection{Computation Flow Abstraction}
\rOneFC{In binary Transformers, weights and activations are in the format of $\alpha x + \beta$, where $\alpha$ and $\beta$ are coefficient and offset under full-precision, and $x$ is a $n$-bit integer (INT) number.}
% \Rone{As a result, full-precision operations are still common in binary Transformer inference.}
% \Rone{Moreover, the speedup potential of binary Transformer cannot be fully explored when performing the basic inference computation flow naively.}
% \Rone{The speedup potential of binary Transformer cannot be fully explored when performing the basic inference computation flow naively because full-precision operations are still common in the process.}
% \Rone{As a result, full-precision operations are still common in binary Transformer inference, and thus the speedup potential of binary Transformer cannot be fully explored when performing the basic inference computation flow naively.}
% \Rone{As a consequence, full-precision operations remain prevalent in binary Transformer inference, and thus the speedup potential of binary Transformer cannot be fully explored when performing the original inference computation flow naively.}
% \Rone{As a consequence, full-precision operations remain prevalent in quantized models inference, which is not hardware-friendly.}
% \fix{When performing multiplication $(\alpha_1 x_1 + \beta_1) \times (\alpha_2 x_2 + \beta_2)$, full-precision operation is executed instead of low-bit operation.}
\refine{When performing multiplication $(\alpha_1 x_1 + \beta_1)\times(\alpha_2 x_2 + \beta_2)$ in Transformer inference on CPU or GPU \cite{BiT, BinaryBERT, BiBERT, BEBERT}, full-precision multiplication is executed instead of integer operation, resulting in heavy energy consumption.}
% \refine{Also, most quantized networks accelerator designs, like \cite{VAQF}, employ quantization method without offset involved, so only offset is considered.}
% \refine{Also, existing quantized Transformer accelerators were not specifically designed for binary Transformers.}
% \refine{While some of them take the coefficients introduced during quantization into account, offsets are not considered\cite{VAQF}.}
% \final{Also, though some of existing quantized Transformer accelerators \cite{VAQF} take the coefficients introduced during quantization into account, offsets are never considered.}
% \final{Also, though some of existing quantized Transformer accelerators \cite{VAQF} take the coefficients introduced during quantization into account, offsets are not considered, which makes them uncompatible with mainstream binary Transformers like BiT \cite{BiT}, BinaryBERT \cite{BinaryBERT} and BiBERT \cite{BiBERT}.}
% \final{Also, existing quantized Transformer accelerators \cite{VAQF} are mostly responsible for deploying quantized Transformers with no consideration on offset $\beta$, which makes them uncompatible with mainstream binary Transformers like BiT \cite{BiT}, BinaryBERT \cite{BinaryBERT} and BiBERT \cite{BiBERT}.}
\final{Also, existing quantized Transformer accelerators are either designed for the deployment of fully quantized Transformers without coefficients and offsets \cite{FQ-BERT, ITA}, or tailored for quantized Transformers that solely consider coefficients without accounting for offsets \cite{VAQF}. This limitation makes them uncompatible with binary Transformers like BiT \cite{BiT}, BinaryBERT \cite{BinaryBERT} and BiBERT \cite{BiBERT}.}
% \fix{If so, the advantages of quantization are not exploited.}
% \refine{Existing algorithmic implementation of binary Transformers  on CPU or GPU,}
% \refine{Existing binary Transformers  designs simply implement full-precision inference on CPU or GPU, with little hardware optimization, resulting in heavy energy waste.}
% \Rone{In order to maximize the advantages of binarization, a computation flow abstraction approach is proposed, which involves adjusting the computation orders to maximize low-bit integer operations and consolidate full-precision numbers, while guaranteeing the correctness of results.}
% \fix{To address this issue, a computation flow abstraction method is developed, which involves adjusting the computation orders and fusing full-precision coefficients and offsets, with no change on the results.}
% \refine{To address this issue, a general computation flow abstraction method is proposed, which involves adjusting the computation orders and fusing full-precision coefficients and offsets to reduce computational complexity, with no change on the results.}
\refine{To fully leverage the energy-efficient potential of binary Transformers, a general computation flow abstraction method is proposed, which involves adjusting the computation orders and fusing full-precision coefficients and offsets to reduce computational complexity without impacting model accuracy.}

\begin{figure}[ht]
    \centering
    \includegraphics[width=0.46\textwidth]{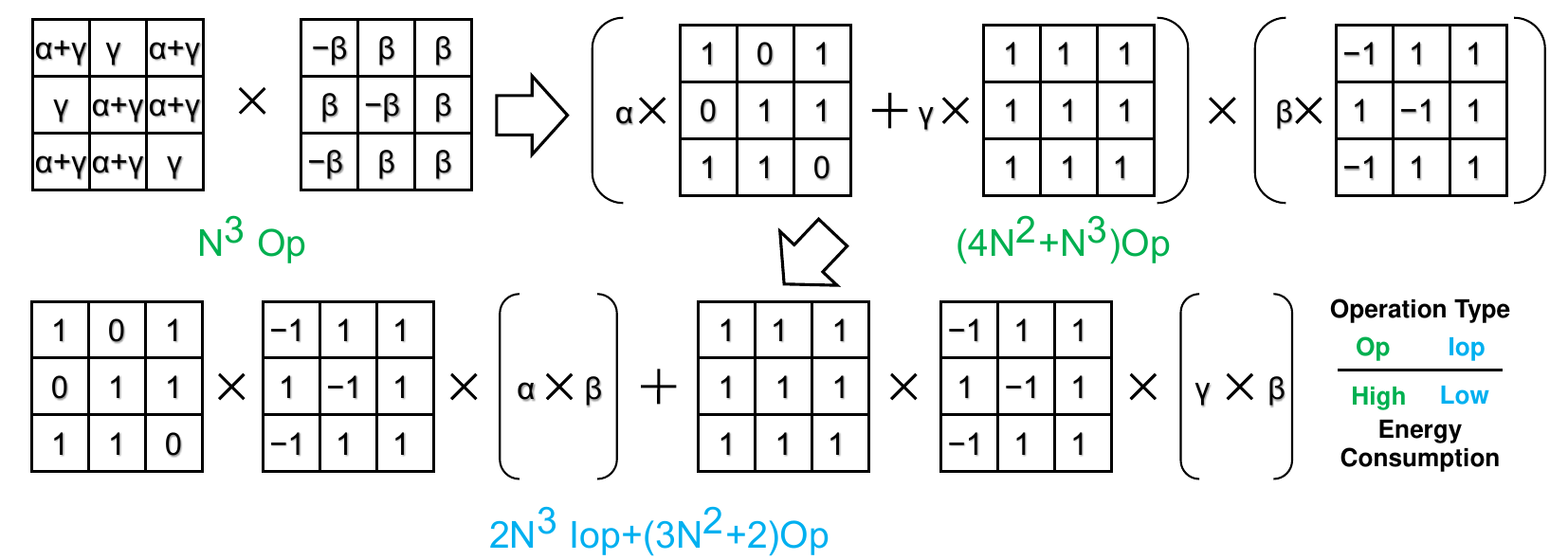}
    % \caption{An example of binary activation$\times$weight operation and its computation flow abstraction process together with corresponding time complexity.}
    \caption{An example of binary activation$\times$weight operation $(\alpha A + \gamma \cdot \mathbf{1}) \times \beta W$ and its computation flow abstraction process together with corresponding \rOneFC{computational} complexity. Full-precision number $\alpha,\beta$ serve as coefficients, $\gamma$ serves as offset, and $A, W$ are binary matrices. Op denotes full-precision operation and Iop denotes integer operation.}
    \label{fig:abstraction}
    \vspace{-0.3cm}
\end{figure}

% \Rone{A simple example of QMM and its abstraction process is illustrated in the Fig.~\ref{fig:abstraction}.}
% \Rone{A binary matrix can be represented as $\alpha X$, where $\alpha \in \mathbb{R}$ serves as coefficient and each element in $X$ is represented as an integer in N-bit width.}
% \fix{Assume a binary matrix can be represented as $\alpha X$ or $\beta Y$, where $\alpha, \beta \in \mathbb{R}$ serves as coefficient and each element in $X, Y$ is 0 or 1.}
% \refine{Assume one binary activation$\times$weight operation is formulated as $(\alpha X + \gamma \cdot \mathbf{1}) \times \beta Y$, as is shown Fig.~\ref{fig:abstraction}, where floating-point number $\alpha,\beta$ serves as coefficients, $\gamma$ serves as offset, and each element in $X, Y$ is 0 or 1.}
\modify{Assume one binary activation$\times$weight operation is formulated as $(\alpha A + \gamma \cdot \mathbf{1}) \times \beta W$, as is shown Fig.~\ref{fig:abstraction}.}
% \Rone{It is worth noting that integers in N bit width can have multiple value ranges, such as $\{0, 1, ..., 2^N-1\}$ or $\{-2^{N-1}, -2^{N-1}+1, ..., 2^{N-1}-1\}$.}
% \Rone{During QMM, leveraging the associativity of multiplication allows to first perform integer matrix operations and consolidate full-precision coefficient.}
% \fix{Instead of simply performing $\alpha X \times \beta Y$, we adjust the multiplication order, turning into $X \times Y \times (\alpha \beta)$.}
\refine{Based on matrix arithmetic, we adjust the computation order, turning the expression into $A \times W \times (\alpha \beta) + \mathbf{1} \times W \times (\gamma \beta)$.}
% \Rone{In this example, time complexity reduces from $N^3$ Flop to $N^3$ Iop $+$ $ (N^2+1)$ Flop after abstraction, where Flop denotes Floating-point operation and Iop denotes Integer operation.}
% \refine{In this case, a floating-point MM is transformed into a combinational operation of integer MM and multiplication by floating-point coefficients, reducing time complexity from $N^3$ Flop to $2N^3$ Iop $+$ $ (2N^2+2)$ Flop, where Flop denotes floating-point operation and Iop denotes integer operation.}
\modify{In this case, a full-precision \final{matrix multiplication (MM)} is transformed into a combinational operation of integer MM and multiplication by full-precision coefficients, reducing time complexity from $N^3$ Op to $2N^3$ Iop $+$ $ (3N^2+2)$ Op.}
\finetune{Noting that both $\alpha\beta$ and $\gamma\beta$ can be performed offline, yielding two new coefficients.}
% \Rone{Given the higher computational efficiency of integer operations compared to floating-point operations, the abstract computation flow significantly improves performance compared to the original inference flow.}
% \Rone{Considering the energy savings of integer operation compared to floating-point operation, which can be several tens or even several hundreds of times, the abstract computation flow significantly improves performance compared to the original inference flow.}
% \Rone{Considering the energy savings of Iop compared to Flop, which can be several tens or even several hundreds of times \cite{DBLP:journals/corr/abs-2306-06446}, the abstract computation flow significantly reduces energy consumption overhead compared to the inference flow $\alpha X \times \beta Y$.}
% \refine{Considering the energy savings of Iop compared to Flop, which can be several tens or even several hundreds of times \cite{DBLP:journals/corr/abs-2306-06446}, the abstract computation flow significantly reduces energy consumption overhead compared to the origin floating-point MM.}
\modify{Considering the energy savings of Iop compared to FP-32 or FIX-16 Ops, which can be several tens or even hundreds of times \cite{DBLP:journals/corr/abs-2306-06446}, the abstract computation flow significantly reduces energy consumption overhead compared to the origin full-precision MM.}
% \modify{Considering the energy savings of Iop compared to Op (FP-32 operation), which can be several tens or even several hundreds of times \cite{DBLP:journals/corr/abs-2306-06446}, the abstract computation flow significantly reduces energy consumption overhead compared to the origin full-precision MM.}
% ref: https://arxiv.org/pdf/2306.06446.pdf
% \update{Other works, like VAQF\cite{VAQF}, include simple computation abstraction, which ignores the weight coefficient temporarily for later process.}
% \Rone{To efficiently support integer matrix multiplication, corresponding hardware component named QMM engine is designed.}

\begin{figure}[ht]
    \centering
    \includegraphics[width=0.48\textwidth]{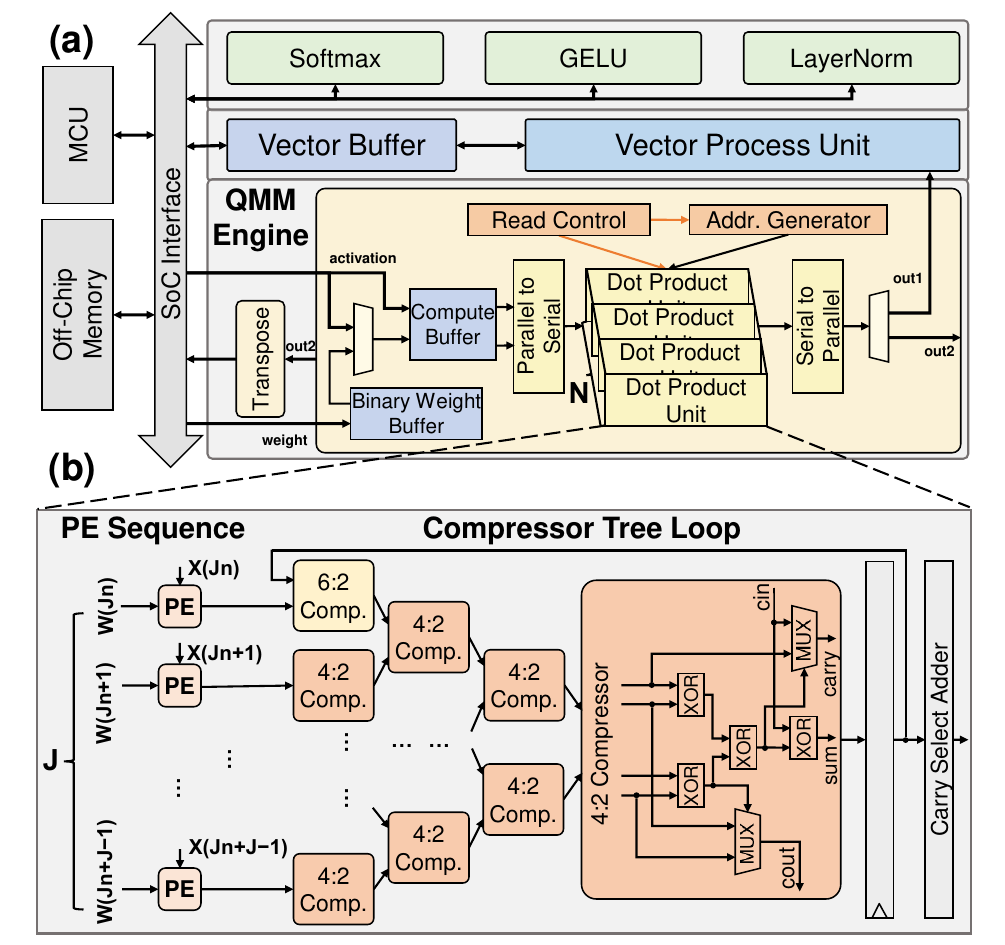}
    % \caption{Overall architecture of BETA. These orange arrows pass control signals, and those black arrows transfer data.}
    % \caption{(a) Hardware overview of BETA, where the orange arrows pass control signals, and the black arrows transfer data. (b) Detailed structure of dot product unit, consisting of the PE sequence and Wallace tree loop.}
    % \caption{(a) Hardware overview of BETA, where the orange arrows pass control signals, and the black arrows transfer data. (b) Detailed structure of dot product unit, consisting of the PE sequence and Wallace tree loop.}
    \caption{(a) Hardware architecture of BETA, where the orange arrows pass control signals, and the black arrows transfer data. (b) Detailed structure of dot product unit, which consists of the PE sequence and \modify{compressor} tree loop.}
    \label{fig:Architecture}
    \vspace{-0.6cm}
\end{figure}

\subsection{Overall Hardware Architecture}
% The proposed PDPU unit is designed based on the architecture depicted in Fig. \ref{fig:architecture}, implemented by a fine-grained 6-stage pipeline. 
% The architecture of the proposed PDPU is depicted in Fig. \ref{fig:architecture}, implemented by a fine-grained 6-stage pipeline.
% \Rone{Fig.~\ref{fig:Architecture} presents the architecture of the proposed BETA consisting of: a QMM engine, a Vector Process Unit (VPU), a Softmax module, a LayerNormalization module, a GELU module, a Micro Control Unit (MCU), the off-chip memory, and several on-chip buffers.}
% \refine{Fig.~\ref{fig:Architecture}~(a) presents the architecture of the proposed BETA consisting of: a QMM engine, a Vector Process Unit (VPU), a Softmax module, a LayerNormalization module, a GELU module, a Micro Control Unit (MCU), the off-chip memory, and several on-chip buffers.}
% \modify{Fig.~\ref{fig:Architecture}~(a) presents the architecture of the proposed BETA consisting of a QMM engine, a Vector Process Unit (VPU), a Micro Control Unit (MCU), the off-chip memory, several non-linear function modules, and on-chip buffers.}
\rOneFC{Fig.~\ref{fig:Architecture}~(a) presents the architecture of the proposed BETA, which comprises a QMM engine, a vector process unit (VPU), several non-linear function modules, and on-chip buffers.}
% \Rone{QMM engine, which consists of $N$ Dot Product Units (DPUs), is used to accelerate quantized matrix multiplication with multi-precision support.}
% \fix{QMM engine, which consists of $N$ Dot Product Units (DPUs), is the core unit, accelerating the key operation QMM with different activation precisions support.}
% \fix{Dominated operations of binary Transformers, i.e., quantized matrix multiplication, are performed by QMM engine with the dynamic configuration under high computational efficiency.}
% \refine{Dominated operations of binary Transformers, i.e. QMM, are performed by QMM engine with dynamic configuration under high computational efficiency.}
% \rOneFC{The dominated operations of binary Transformers, i.e. QMM in the binary linear layers, are performed by QMM engine with dynamic configuration to ensure high computational efficiency.}
% \rOneFC{The dominated operations of binary Transformers, \final{QMMs}, are performed by QMM engine with dynamic configuration to ensure high computational efficiency.}
\finetune{QMMs, the dominant operations of binary Transformers, are performed by QMM engine with dynamic configuration and high computational efficiency.}
% \Rone{VPU takes charge of implementing full-precision operations involved in the abstract computation flow including coefficient multiplication, and offset addition, with vectorial inputs and outputs.}
% \Rone{VPU takes charge of implementing full-precision operations involved in the abstract computation flow including coefficient multiplication, and offset addition, with \modify{vectorized} inputs and outputs.}
\finetune{VPU is responsible for the implementation of full-precision operations involved in the abstract computation flow including coefficient multiplication, and offset addition, with vectorized inputs and outputs.}
% \Rone{The Softmax module, LayerNormalization module and GELU module are responsible for the non-linear functions corresponding to Softmax, LayerNormalization and GELU.}
% \update{Since non-linear functions, Softmax, Layer Normalization and GELU, are not so compute-intensive as QMM, their parameters and inputs are kept with full-precision on the corresponding hardware modules to preserve the model accuracy.}
% \rOneFC{As non-linear functions, including Softmax, layer normalization, and GELU, are not as computationally intensive as QMM, their operations are maintained with full precision on the corresponding modules to preserve the model accuracy.}
% \rOneFC{As non-linear functions, including Softmax, \final{Layer Normalization}, and GELU, are not as computationally intensive as QMM, their operations are maintained with full precision on the corresponding modules to preserve the model accuracy.}
\publish{As non-linear functions, including Softmax, Layer Normalization, and GELU, are not as computationally intensive as QMM, their operations are maintained with full precision to preserve the model accuracy.}
% \Rone{Since different networks may employ different quantization methods in which complex operations such as division, rounding, and square root are involved, it is challenging to devise a single quantization unit that can effectively handle various quantization schemes.}
% \Rone{So a MCU is used to perform the quantization methods needed.}
% \Rone{A MCU is used to perform the quantization functions.}
% \update{The host MCU is used to perform quantization functions, which occupies small latency overhead for embedded FPGAs.}
\rOneFC{The host MCU is utilized for quantization functions, incurring minimal latency overhead for the inference of binary Transformers.}
% \Rone{Also, to make full use of the advantage of binarization, weights are stored on the on-chip buffer, which significantly eases the off-chip data access pressure.}
% \Rone{Also, considering the limited parameter size of binarization, weights are stored on the on-chip buffer, which significantly eases the off-chip data access pressure.}
% \fix{Considering the limited parameter size of binarization, the whole matrices under computation are loaded to the compute buffer before performing QMM, which significantly improves data reuse and reduces data access bandwidth needed compared to loading partial matrices from off-chip memory each time.}
% \refine{Due to the limited parameter size of binarization, the entire matrices involved in the computation are loaded into the compute buffer from off-chip memory before performing QMM, which substantially enhances data reuse and minimizes the required data access bandwidth compared to loading partial matrices each time.}
% % \fix{Also, binary weights are stored on the on-chip buffer.}
% \fix{Weight buffer are used to store binary model weights.}

\subsection{QMM engine}
\final{To improve the overall hardware efficiency and support different types of QMM, QMM engine is designed with a focus on high throughput and configurability.}
% \Rone{Fig.~\ref{fig:Architecture} also presents the detailed architecture of QMM engine.}
% \refine{Fig.~\ref{fig:Architecture}~(a) depicts the architecture of QMM engine, which consists of $N$ dot product units (DPUs), a compute buffer, a binary weight buffer, and various control logics.}
\rOneFC{As shown in Fig.~\ref{fig:Architecture}~(a), it consists of $N$-parallel dot product units (DPUs), a compute buffer, a binary weight buffer, and various control logics.}
% \refine{Due to the limited parameter size of binarization, the entire matrices involved in the computation are loaded into the compute buffer from off-chip memory before performing QMM, which substantially enhances data reuse and minimizes the required data access bandwidth compared to loading partial matrices each time.}
% \fix{Binary weight buffer are used to store all the binary model weights before computation.}
% \modify{Due to the limited parameter size of binarization, all binary weights are stored in the on-chip binary weight buffer before inference.}
% \modify{Due to the small size of binary models, all binary weights in the whole model are stored in the on-chip binary weight buffer before inference.}
% \modify{Binary weights are stored in the on-chip binary weight buffer before inference.}
\publish{Binary weights are stored in the on-chip buffer before inference.}
% \modify{When performing QMM, the entire matrices involved in the computation are pre-loaded to the compute buffer from off-chip memory or weight buffer.}
% \modify{These memory scheduling methods enhance data reuse and minimize the required data access bandwidth.}
\modify{When performing QMM, the entire matrices involved in the computation are pre-loaded to the compute buffer from off-chip memory or weight buffer, \publish{which enhances data reuse and minimizes the required data access bandwidth}.}
\refine{Fig.~\ref{fig:Architecture}~(b) shows the detailed structure of DPU.}
% \refine{Each DPU is composed of a PE sequence and a Wallace tree loop.}
\refine{Each DPU is composed of a PE sequence and a \modify{compressor} tree loop.}
% \refine{Besides duplicating DPU for $N$ times, processing different dot product operations simultaneously, we further explore the parallelism within one vector using unfolding method, which means $J$ elements in one vector are processed at one time.}
% \refine{Besides duplicating DPU for $N$ times to process $N$ dot product operations simultaneously, we further explore the parallelism within one vector using unfolding method, which means $J$ elements in one vector are processed each time.}
% \refine{Besides duplicating DPU for $N$ times to process $N$ dot product operations simultaneously, we further explore the parallelism within one vector using unfolding \modify{skill}, \modify{in which} $J$ elements in one vector are processed each time.}
\rOneFC{In addition to replicating DPUs for $N$ times to process dot product operations simultaneously, we further leverage unfolding techniques to exploit parallelism within a single vector.}
\rOneFC{A DPU can process $J$ elements from one vector at a time after unfolding.}
\rOneFC{Both replication and unfolding techniques increase the parallelism of QMM engine. Note that the factor of $N$ and $J$ can be flexibly adjusted based on the desired level of parallelism.}
\final{To reduce the circuit delay of unfolding structure, we design a compressor tree loop for dot product accumulation.}
% \update{Wallace tree structure is utilized to compress the $J$ computed results and two accumulation results, thereby causing the growth of the loop delay to limited logarithmic relationship with $J$, as is illustrated in Fig.~\ref{fig:Architecture}~(b).}
% \modify{Compressor tree is built to compress the $J$ computed results and two accumulation partial results, thereby causing the growth of the loop delay to limited logarithmic relationship with $J$, as is illustrated in Fig.~\ref{fig:Architecture}~(b).}
\final{Compressor-based adder tree is built to aggregate the $J$ computed results and two accumulation partial results, thereby mitigating the carry chain propagation and limiting the loop delay to logarithmic relationship with $J$, as is illustrated in Fig.~\ref{fig:Architecture}~(b).}
\final{The two outputs of compressor tree loop are sent to a carry select adder to generate the final result of dot product.}
\refine{The parallelism improvement and circuit delay reduction result in high throughput.}
% \Rone{QMM engine is able to perform various types of QMM, such as $Q \times K$, $S \times V$, etc., in various types of binary Transformer networks such as BinaryBERT, BiBERT and BiT, in different activation quantization precisions such as 1-bit, 2-bit, 4-bit, etc..}
% \Rone{As is illustrated in Fig[], PE is designed with high scalability, where different computation modes are supported.}
% % 解释mode是什么
% \Rone{As such, it is able to perform various types of QMM, such as $Q \times K$, $S \times V$, etc., in various binary Transformer networks such as BinaryBERT, BiBERT and BiT, in different activation quantization precisions such as 1-bit, 2-bit, 4-bit, etc..}

\begin{figure}[ht]
    \centering
    \includegraphics[width=0.44\textwidth]{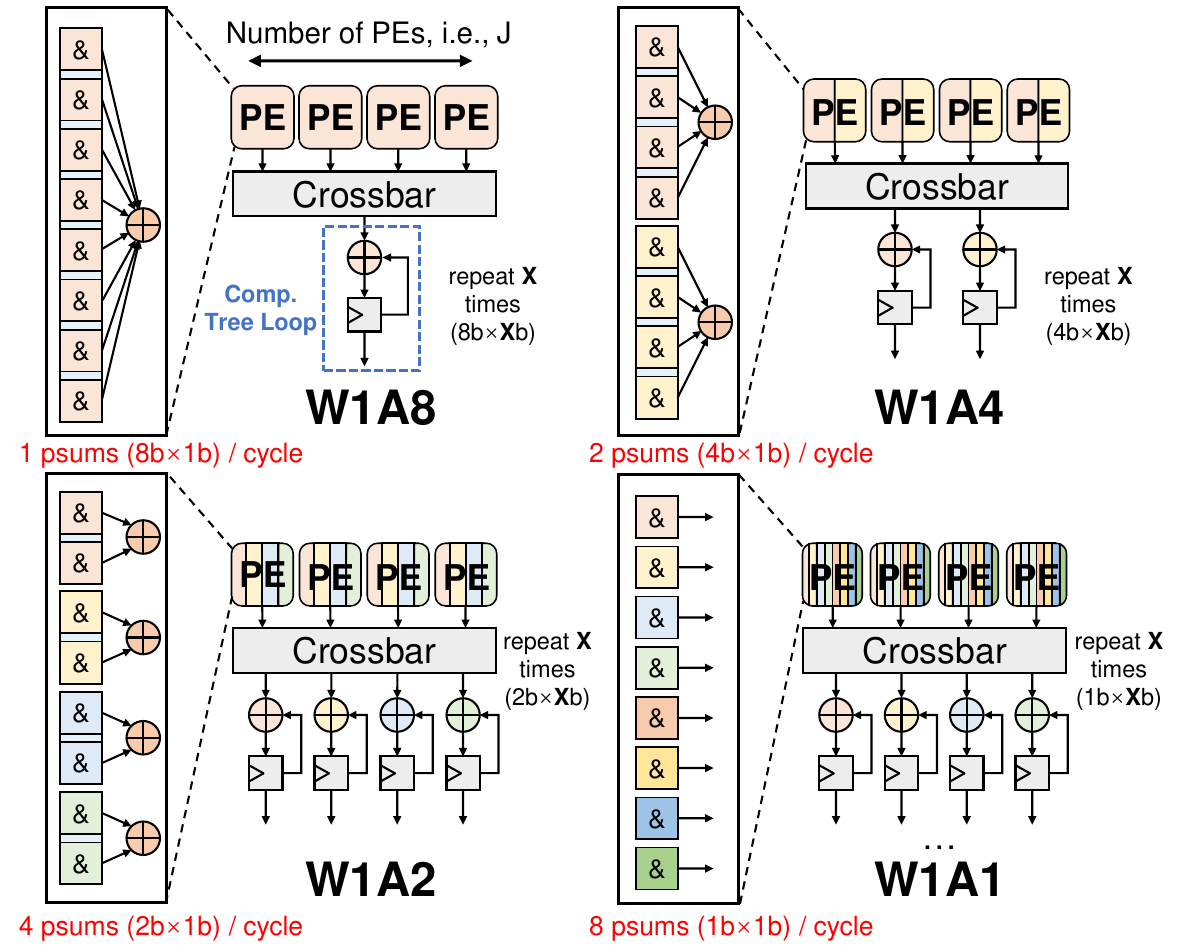}
    % \caption{Hierarchical architecture of PE. It consists of a Configurable Multiplier and several computation blocks. The detailed structure of Configurable Multiplier is shown with two modes on each level, compute mode (top) and concat mode (bottom).}
    % \caption{Hierarchical structure of PE. It consists of a bit-level configurable multiplier and a shift-add block, which perform different operations according to activation precision and QMM types.}
    % \caption{Operation modes of configurable PE sequence, which adopts data-packing and bit-serial to enable flexible configuration to process different inputs.}
    \caption{\Rone{Operation modes of configurable PE sequence, which combines data-packing and bit-serial to enable flexible configuration to process different workload.} \update{Note that a network with weights quantized to $b_w$ bits and activations quantized to $b_a$ bits is denoted as $Wb_wAb_a$ \cite{BiT}.}}
    \label{fig:pe}
    \vspace{-0.6cm}
\end{figure}

\rOneFC{As shown in Fig.~\ref{fig:pe}, PE sequence in DPU can flexibly perform computations by configuring packing format and accumulation times according to the combination of activation precision and QMM type.}
\final{For example, when performing binary weight $\times$ 4-bit activation in W1A4 mode, two multiplications are executed simultaneously and the results are packed in 8-bit output of one PE, with one cycle needed.}
% \update{Furthermore, when the QMM type is 4-bit activation$\times$activation, bit-level serial is achieved by traversing one operand within four cycles.}
\final{Furthermore, when the QMM type is 4-bit activation$\times$activation, one input operand is traversed on bit-level within four cycles to generate the results.}
\renewcommand\arraystretch{1.0}
\begin{table}[h]
\centering
\caption{FPGA resource breakdown of BETA}
\label{tab:resource}
\resizebox{\columnwidth}{!}{%
\begin{tabular}{cc|cccc}
\hline
 &  & LUT & FF & BRAM & DSP \\ \hline
\multicolumn{1}{c|}{\multirow{4}{*}{QMM Engine}} & \begin{tabular}[c]{@{}c@{}}Dot Product \\ Unit\end{tabular} & 154K & 49K & - & - \\ \cline{2-6} 
\multicolumn{1}{c|}{} & \begin{tabular}[c]{@{}c@{}}Compute\&Weight \\ Buffer\end{tabular} & - & - & 456 &  \\ \cline{2-6} 
\multicolumn{1}{c|}{} & Others & 21K & 25K & - & - \\ \hline
\multicolumn{2}{c|}{VPU} & 4K & - & - & 64 \\ \hline
\multicolumn{2}{c|}{Others} & 12K & 14K & 87 & - \\ \hline
\multicolumn{2}{c|}{Total} & 191K & 88K & 543 & 64 \\ \hline
\multicolumn{2}{c|}{Utilization} & \begin{tabular}[c]{@{}c@{}}274K\\ (69.71\%)\end{tabular} & \begin{tabular}[c]{@{}c@{}}548K\\ (16.06\%)\end{tabular} & \begin{tabular}[c]{@{}c@{}}912\\ (59.54\%)\end{tabular} & \begin{tabular}[c]{@{}c@{}}2520\\ (2.54\%)\end{tabular} \\ \hline
\end{tabular}%
}
\vspace{-0.4cm}
\end{table}

\section{Experimental Results}\label{sec:results}
\renewcommand\arraystretch{1.5}
\begin{table*}
\centering
\caption{Comparison of BETA with previous works and commercial products}
\label{tab:comparison}
\resizebox{\textwidth}{!}{%
\begin{threeparttable}
\begin{tabular}{c|cllcll|ccccccllccc}
\hline
\multirow{4}{*}{Platform} & \multicolumn{3}{c}{\multirow{3}{*}{CPU}} & \multicolumn{3}{c|}{\multirow{3}{*}{GPU}} & \multicolumn{11}{c}{FPGA} \\ \cline{8-18} 
 & \multicolumn{3}{c}{} & \multicolumn{3}{c|}{} & \multirow{2}{*}{ViA\cite{ViA}} & \multirow{2}{*}{STA\cite{STA}} & \multirow{2}{*}{EFA-Trans\cite{EFA-Trans}} & \multicolumn{1}{c|}{\multirow{2}{*}{VAQF\cite{VAQF}}} & \multicolumn{7}{c}{Our Work} \\ \cline{12-18} 
 & \multicolumn{3}{c}{} & \multicolumn{3}{c|}{} &  &  &  & \multicolumn{1}{c|}{} & Baseline1 & \multicolumn{3}{c|}{Baseline2} & \multicolumn{3}{c}{BETA} \\ \cline{2-18} 
 & \multicolumn{3}{c}{i7-10510U} & \multicolumn{3}{c|}{RTX 3090} & Alveo U50 & ZC702 & ZCU102 & \multicolumn{1}{c|}{ZCU102} & \multicolumn{7}{c}{ZCU102} \\ \hline
Technology & \multicolumn{3}{c}{14nm} & \multicolumn{3}{c|}{8nm} & 16nm & 16nm & 16nm & \multicolumn{1}{c|}{16nm} & \multicolumn{7}{c}{16nm} \\ \hline
Frequency (Hz) & \multicolumn{3}{c}{1.8G} & \multicolumn{3}{c|}{1.7G} & 300M & 150M & N/A & \multicolumn{1}{c|}{150M} & \multicolumn{7}{c}{190M} \\ \hline
Test Network & \multicolumn{6}{c|}{BiT} & \begin{tabular}[c]{@{}c@{}}Swin\\ Transformer\end{tabular} & \begin{tabular}[c]{@{}c@{}}N:M Sparse\\ Transformer\end{tabular} & \begin{tabular}[c]{@{}c@{}}Sparse\\ Transformer\end{tabular} & \multicolumn{1}{c|}{\begin{tabular}[c]{@{}c@{}}Quantized Vision\\ Transformer\end{tabular}} & \multicolumn{4}{c|}{BiT} & BiT & BinaryBERT & BiBERT \\ \hline
Computation Abstraction & \multicolumn{3}{c}{\XSolidBrush} & \multicolumn{3}{c|}{\XSolidBrush} & \XSolidBrush & \XSolidBrush & \XSolidBrush & \multicolumn{1}{c|}{\CheckmarkBold} & \XSolidBrush & \multicolumn{3}{c|}{\XSolidBrush} & \CheckmarkBold & \CheckmarkBold & \CheckmarkBold \\ \hline
BiT Precision & \multicolumn{3}{c}{FP-32} & \multicolumn{3}{c|}{FP-32} & FP-16 & FIX-16 & FIX-8 & \multicolumn{1}{c|}{INT\&FIX-16\tnote{$\dagger$}} & FP-32 & \multicolumn{3}{c|}{FIX-16} & \multicolumn{3}{c}{INT\&FIX-16\tnote{$\dagger$}} \\ \hline
W/A Precision & \multicolumn{3}{c}{W1A1} & \multicolumn{3}{c|}{W1A1} & W16A16 & W16A16 & W8A8 & \multicolumn{1}{c|}{W1A8} & W1A1 & \multicolumn{3}{c|}{W1A1} & \multicolumn{3}{c}{W1A1} \\ \hline
Throughput (GOPS) & \multicolumn{3}{c}{6.69} & \multicolumn{3}{c|}{484.26} & 309.60 & 109.45 & 279.80 & \multicolumn{1}{c|}{861.20} & 13.51 & \multicolumn{3}{c|}{72.09} & 1240.98 & 1387.59 & 1436.07 \\ \hline
Power (W) & \multicolumn{3}{c}{25.00} & \multicolumn{3}{c|}{350.00} & 38.99 & 2.71 & 5.48 & \multicolumn{1}{c|}{8.70} & 11.64 & \multicolumn{3}{c|}{3.91} & 7.18 & 7.95 & 8.20 \\ \hline
Energy Efficiency (GOPS/W) & \multicolumn{3}{c}{0.27} & \multicolumn{3}{c|}{1.38} & 7.94 & 40.39 & 51.06 & \multicolumn{1}{c|}{98.98} & 1.16 & \multicolumn{3}{c|}{18.42} & 172.41 & 174.59 & 175.23 \\ \hline
\end{tabular}%
\begin{tablenotes}
    % \item[$\dagger$] In abstract computation flow, according to Fig.~\ref{fig:abstraction}, integer (INT) MM and full-precision coefficient multiplications and offset additions are executed. FIX-16 numbers are employed to perform full-precision operations. 
    % \item[$\dagger$] \finetune{According to the abstract computation flow in Fig.~\ref{fig:abstraction}, integer (INT) operations are performed in QMMs, and FIX-16 format is used as full-precision to perform coefficient multiplications and offset additions.}
    \item[$\dagger$] \finetune{According to the abstract computation flow in Fig.~\ref{fig:abstraction}, BETA performs integer (INT) operations in QMMs. And here FIX-16 format is used as full-precision to perform coefficient multiplications and offset additions.}
\end{tablenotes}
\end{threeparttable}
}
\vspace{-0.5cm}
\end{table*}
\subsection{Experimental Setup}
% \Rone{BETA with different precisions support is implemented using SystemVerilog on the Xilinx XCZU9EG FPGA platform with 274k LUTs, 548k FFs, 912 BRAMs and 2520 DSPs.}
% \Rone{BETA with different precisions support is implemented using SystemVerilog on the Xilinx XCZU9EG FPGA platform with 274k LUTs and 912 BRAMs.}
% \Rone{Vivado 2022.2 is the tool for synthesis and implementation.}
% \Rone{To maximize the computation efficiency and avoid timing violation, the frequency is set to 210MHz.}
% \Rone{The size of input tensor is $128 \times 512$ with the corresponding parallelism configuration is $N=2$ and $J=256$.}
% \Rone{We implement the complete design of our BETA architecture in SystemVerilog and evaluate the matrix multiplication relative part for the BiT, BinaryBERT and BiBERT architecture on the Xilinx XCZU9EG FPGA platform with 274k LUTs and 912 BRAMs.}
% \fix{We implement the complete design of our BETA architecture in SystemVerilog and evaluate the matrix multiplication relative part for BiT, BinaryBERT and BiBERT on the Xilinx ZCU102 FPGA platform with 274k LUTs and 912 BRAMs.}
% \modify{We implement the complete design of our BETA architecture in SystemVerilog and evaluate the matrix multiplication relative part for BiT, BinaryBERT and BiBERT on the Xilinx ZCU102 FPGA platform.}
\rOneFC{BETA is implemented using Vivado 2022.2 on the Xilinx ZCU102 FPGA platform and evaluated under the benchmarks embracing state-of-the-art binary Transformers \cite{BiT, BiBERT, BinaryBERT}.}
% \Rone{Vivado 2022.2 is the tool for synthesis, implementation, timing violation checking, and power analysis.}
% \Rone{We conduct simulation using Mentor Graphic Modelsim 2020.4 and the generated .vcd files are sent to Synopsys Design Compiler to translate to .saif files for more accurate power analysis.}
\rOneFC{We conduct the functional simulation with the extracted actual data from benchmarks and measure the inference latency of BETA. Meanwhile, we generate the annotated toggle rate from the simulator and dump it into the switching activity interchange format (SAIF). Then, power consumption is estimated by incorporating the SAIF file into the Vivado Power Analysis Tool.}
\Rone{Moreover, for cross-platform comparison, we perform the inference of the benchmark models on an Intel i7-10510U CPU and an NVIDIA RTX 3090 GPU, respectively.}
\rOneFC{Note that data at the edge are usually mini-batch, and therefore the cross-platform comparison is performed with a single batch size on various platforms.}

\subsection{Hardware Consumption}
% \Rone{To maximize the computation efficiency and avoid timing violation, the frequency is set to 190MHz.}
% \Rone{The configuration of parallelism is $N$=2 and $J$=256.}
% \Rone{Table \ref{tab:resource} presents the FPGA resource and average power breakdown of QMM Engine and VPU.}
\rOneFC{The running frequency of BETA is 190MHz, and the configuration of parallelism is $N$=2 and $J$=256.}
\update{Table \ref{tab:resource} presents the FPGA resource breakdown of BETA.}
% \Rone{Since binary Transformers have similar architecture, the power when running different networks are almost the same.}
% \Rone{The dot-product units and Wallace Tree loops dominate the LUT consumption since it is the computing core of QMM.}
% \fix{Dot product units, including PE sequences and Wallace Tree loops, dominate the LUT consumption since they are the computing core of QMM.}
% \fix{\modify{DPUs}, including PE sequences and Wallace Tree loops, dominate the LUT consumption since they are the computing core of QMM.}
\fix{\modify{DPUs}, including PE sequences and \modify{compressor} tree loops, dominate the LUT consumption since they are the computing core of QMM.}
\Rone{Most of BRAMs are occupied by compute buffer and weight buffer to store inputs and binary weights, respectively.}
% \Rone{The VPU uses DSP to perform high-speed fixed-point operations.}
% \Rone{The VPU uses DSP to perform high-speed \modify{FIX-16} operations.}

% \Rone{To validate the advantage of binarization, the input length and hidden size are set to 128 and 512 respectively with all the weights stored on buffer.}
% \subsection{Comparison with the State-of-the-arts}

\begin{figure}[tbp]
    \centering
    \includegraphics[width=\linewidth]{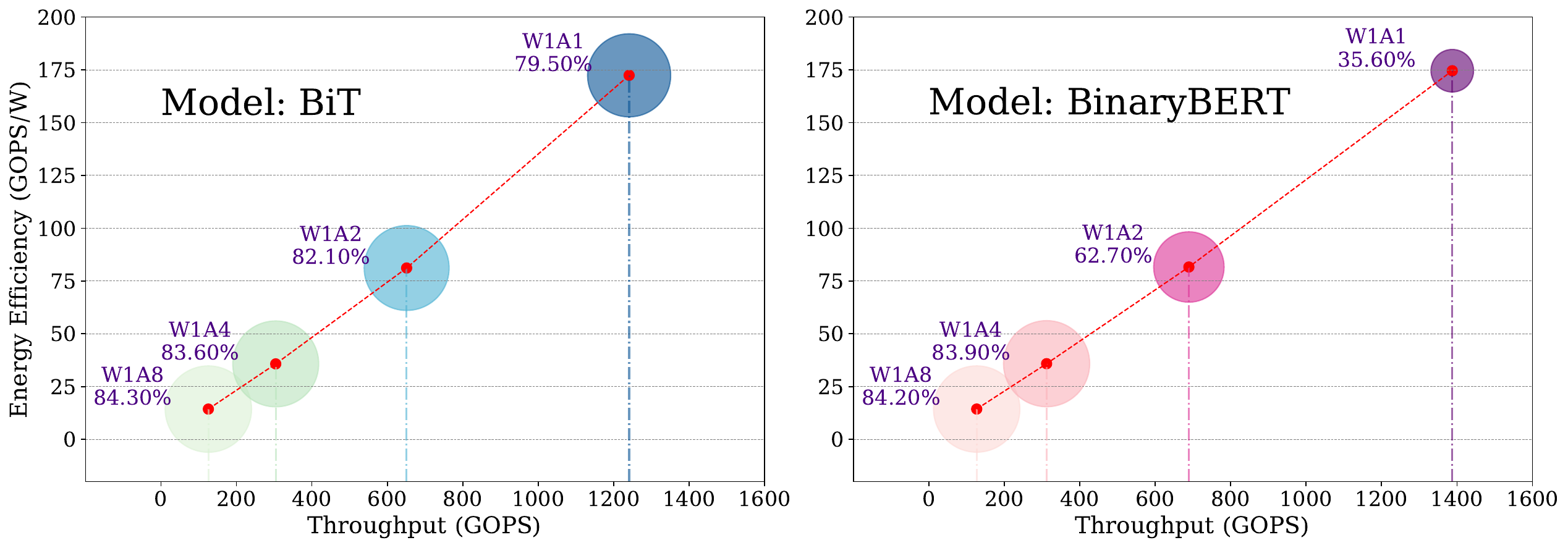}
    \caption{\Rone{Tradeoff between hardware efficiency and model accuracy on BETA.}}
    \label{fig:dyna-adjustment}
    \vspace{-0.5cm}
\end{figure}

\subsection{Dynamic Adjustment between Efficiency and Accuracy}
\rOneFC{We evaluate BiT and BinaryBERT with different activation precisions on BETA, collecting throughput, energy efficiency, and model accuracy on the MNLI-m dataset \cite{GLUE} to \finetune{understand} the tradeoff between hardware efficiency and model accuracy.}
% \Rone{From W1A8 to W1A1, the energy efficiency and throughput both improves about 8x for BiT and BinaryBERT as we expected, with the accuracy loss of 4.7\% and 48.6\%, respectively, which demonstrates the ability of BETA to dynamically adjust hardware efficiency and model accuracy.}
% \fix{As is presented in Fig.~\ref{fig:dyna-adjustment}, with activation precision decreasing, the energy efficiency and throughput both improve obviously for BiT and BinaryBERT, while the model accuracy drops.}
% \update{As shown in Fig.~\ref{fig:dyna-adjustment}, as activation precision decreases, the energy efficiency and throughput both improve obviously for BiT and BinaryBERT, while the model accuracy drops.}
% \update{As shown in Fig.~\ref{fig:dyna-adjustment}, \modify{when} activation precision decreases, the energy efficiency and throughput both improve obviously for BiT and BinaryBERT, while the model accuracy drops.}
% \modify{As shown in Fig.~\ref{fig:dyna-adjustment}, \modify{when} activation precision decreases, the energy efficiency and throughput both improve obviously.}
\final{As shown in Fig.~\ref{fig:dyna-adjustment}, when the activation precision of the deployed model decreases, there is a notable improvement in both throughput and energy efficiency on BETA, while conversely, the model accuracy gradually drops.}
% \final{For BiT, the energy efficiency and throughput improves 12.02$\times$ and 9.88$\times$, respectively, while the model accuracy drops from 84.3\% to 79.5\% from W1A8 to W1A1 version.}
% \Rone{However, the model accuracy drops from 84.3\% to 79.5\% for BiT and 84.2\% to 35.6\% for BinaryBERT.}
% \Rone{BiBERT only has W1A1 version with accuracy of 67.3\%.}
% \refine{BiBERT only has W1A1 version, which presents the highest throughput with accuracy of 67.3\%.}
% \modify{Noting that BiBERT only has W1A1 version with accuracy of 67.3\%.}
% \modify{Noting that BiBERT only has W1A1 version.}
% \update{This experiment validates the flexibility of BETA to dynamically adjust hardware efficiency and model accuracy, which is suitable for multiple edge deployment scenarios.}
% \final{This experiment proves that BETA is suitable for multiple edge deployment scenarios due to its configurability.}
\final{This experiment demonstrates that BETA enables dynamic adjustment between model inference efficiency and accuracy, which allows it to meet deployment requirements in various edge scenarios with different constraints.}
% \Rone{Moreover, the results show BETA has a comparable or better performance with previous works on quantized or sparse Transformer-based networks accelerator design.}

\subsection{Comparison with Baselines and Other Architectures}
\rOneFC{We first compare BETA with FP-32 and FIX-16 baselines. Both baselines are implemented on the same FPGA as BETA with \finetune{close} resource consumption, but use traditional FP-32 or FIX-16 computing units instead of BETA's computation flow abstraction DPUs.}
% \Rone{For a fair comparison, the proposed two baselines: FP-32 dot-product unit using floating-point IP core generated by Vivado and FIX-16 dot-product unit both perform full-precision inference of unabstract BiT on the same FPGA as BETA with close resources consumption.}
% \Rone{Compared with two baselines, BETA achieves [] and [] improvement on throughput and [] and [] improvement on energy efficiency, respectively, which significantly demonstrates the necessity of computation flow abstraction.}
% \update{As shown in Table \ref{tab:comparison}, compared with two baselines, BETA achieves 66.94x and 17.21x improvement on throughput and 282.63x and 9.36x improvement on energy efficiency, respectively, which significantly demonstrates the necessity of computation flow abstraction.}
% \refine{As shown in Table \ref{tab:comparison}, compared with FP-32 dot-product unit and FIX-16 dot-product unit, BETA achieves 66.94x and 17.21x improvement on throughput and 282.64x and 9.36x improvement on energy efficiency, respectively, which demonstrates the necessity of computation flow abstraction.}
% \refine{As shown in Table \ref{tab:comparison}, compared with FP-32 dot-product unit and FIX-16 dot-product unit, BETA achieves \modify{91.86}\modify{$\times$} and 17.21\modify{$\times$} improvement on throughput and \modify{148.63}\modify{$\times$} and 9.36\modify{$\times$} improvement on energy efficiency, respectively, which demonstrates the necessity of computation flow abstraction.}
\refine{As shown in Table \ref{tab:comparison}, compared with FP-32 and FIX-16 baselines, BETA exhibits \modify{91.86}\modify{$\times$} and 17.21\modify{$\times$} improvement on throughput and \modify{148.63}\modify{$\times$} and 9.36\modify{$\times$} improvement on energy efficiency, respectively.}
% \Rone{To validate the necessity of computation flow abstraction, we compare BETA with two baselines: FP-32 dot-product unit using floating-point IP core generated by Vivado and FIX-16 dot-product unit, both to perform full-precision inference of unabstract binary Transformers.}

\Rone{Moreover, we compare BETA with other previous FPGA-based works and commercial CPU and GPU products.}
% \Rone{In Table \ref{tab:Comparison}, we list some related works that design accelerator for transformer.}
% \Rone{As illustrated, BETA achieves a significantly higher performance comparing with CPU and GPU due to the superior advantages of binarization.}
% \update{VAQF only performs 1-bit weight $\times$ 8-bit activation operation without 8-bit $\times$ 8-bit activation, resulting in excellent data, however, the real W1A8 performance should be much worse.}
% \fix{VAQF only performs 1-bit weight $\times$ 8-bit activation operation without 8-bit $\times$ 8-bit activation, resulting in unreal W8A8 data, which is hard to be compared.}
% \update{W1A8 Transformer block includes two 8-bit activations multiplication, such as query $\times$ key in self-attention, which is not considered in VAQF design, resulting in unreal W1A8 data.}
% \refine{VAQF\cite{VAQF} fully exploits the speedup potential of binarization, turns multiplication to BiT-wise operation, and presents excellent data results.}
% \modify{VAQF\cite{VAQF} fully exploits the speedup potential of binarization by turning multiplication to BiT-wise operation.}
\final{VAQF\cite{VAQF} turns multiplication involved in MM to bit-wise operation and presents excellent experimental results.}
% \refine{Compared to it, our BETA further supports multi-precision activations multiplication, such as 8-bit query$\times$key in W1A8 self-attention, realizing the true implementation of binary Transformers inference.}
% \refine{Compared to it, our BETA further supports multi-precision activations multiplication, such as 8-bit query$\times$key in W1A8 self-attention, \modify{achieving} the true implementation of binary Transformers inference.}
% \final{Compared to VAQF, BETA further improves the versatility, supporting multi-precision activation$\times$activation operations in a unified computation engine, such as 8-bit query$\times$key in W1A8 self-attention, and achieves 1.76$\times$ energy efficiency improvement.}
% \final{Compared to it, BETA further supports multi-precision activation$\times$activation operations in a unified computation engine, such as 8-bit query$\times$key in W1A8 self-attention, and achieves 1.76$\times$ energy efficiency improvement.}
\final{In contrast, BETA further supports multi-precision activation$\times$activation operations in a unified computation engine, such as 8-bit query$\times$key in W1A8 self-attention.}
% \update{Also, VAQF can only support one activation version for each compilation}
% \fix{Our BETA owns great scalability, while VAQF can only support one activation version for each compilation.}
% \update{Our BETA fully support all kinds of QMM together with precision scalability, while VAQF can only support one precision version for each compilation.}
% \refine{Also, our BETA supports all kinds of QMM together with precision configurability in one unified computation engine, while VAQF can only support one precision version for each compilation.}
% \refine{Also, BETA supports all \modify{types} of QMM together with precision configurability in one unified computation engine, while VAQF can only support one precision version for each compilation.}
% \refine{Also, BETA supports all \modify{types} of QMM with precision configurability in a unified computation engine, \modify{while VAQF generates one precision version hardware in each compilation.}}
% \fix{ViA accelerates full-precision Transformer without quantization with relatively low performance, revealing the advantage of model compression.}
% \update{ViA accelerates full-precision Transformer without quantization with relatively low performance.}
% \refine{ViA\cite{ViA} optimizes hardware architecture based on the characteristics of Vision Transformer (ViT), however, FP-16 networks are processed without quantization, resulting in much more energy consumption relative to our low-bit design.}
\refine{ViA\cite{ViA} deploys FP-16 networks without quantization, resulting in much more energy consumption relative to our low-bit design.}
\final{STA \cite{STA} and EFA-Trans \cite{EFA-Trans} are both dedicated on deploying another kind of compressed Transformers, namely sparse Transformers, and also achieve considerable hardware performance.}
\final{Compared to the FPGA-based accelerators mentioned above, BETA presents 1.76$\sim$21.92$\times$ higher energy efficiency improvement.}
% \Rone{In addition, compared to CPU and GPU, BETA achieves 182.50\modify{$\times$} and \modify{2.56}\modify{$\times$} speedup, respectively.}
\modify{In addition, compared to CPU and GPU, BETA achieves 643.32\modify{$\times$} and \modify{124.93}\modify{$\times$} energy efficiency improvement, respectively.}

% Exp1: Resource Breakdown：单独一小节？讲什么？说明什么
% 首先说明实验平台（资源限制），综合工具，频率，再说明并行度的配置，比如N=2，J=256。说明各部分breakdown的意义，unfolding系数大，绝大多数的LUT和FF花在了DPU（PE和Compressor Loop）；片上weight buffer主要利用BRAM实现；VPU进行定点向量运算时，占用较多的DSP。总功耗是多少，能量开销集中在PE上，或者是静态功耗
% Exp2：Accuracy和能效的动态调整
% 说明在MNLI-m数据集上的精度，BiBERT只有W1A1的情况，说明BiT和BinaryBERT随着激活比特降低，能效比提高，但精度下降的情况，指出适应边缘端场景
% Exp3：出色的模型性能
% 和baseline1,2比较；和C其他FPGA工作比较；和CPU，GPU比较，尽可能解释优点以及原因
% Exp23交换一下顺序，更好的和contribution对应
% (?<=\{c\|\}\{.*\}) 匹配所有{c|}{ABC}后面的位置

% \input{1-intro-r1.tex}
% \input{2-bkg-r1.tex}
% \input{3-design.tex}
% \input{4-results.tex}
\section{Conclusion}\label{sec:concls}
% \Rone{In this paper, we first propose a customized computation flow abstraction method for binary Transformers, which significantly reduce the computation complexity.}
% \refine{In this paper, we first propose a general computation flow abstraction method for binary Transformers, which significantly reduce the computation complexity.}
% \modify{In this paper, we first propose a computation flow abstraction method for binary Transformers, which significantly reduce the computation complexity.}
\rOneFC{In this paper, we develop a computation flow abstraction method and propose a binary Transformer accelerator called BETA to enable flexible and effcient deployment of binarized Transformers at the edge.}
\rOneFC{BETA features a configurable quantized matrix multiplication (QMM) engine that supports diverse activation precisions and offers high parallelism and speed for QMMs with impressive energy efficiency. Experimental results show that BETA achieves an average energy efficiency of 174 GOPS/W, which is 1.76$\sim$21.92$\times$ higher than prior FPGA-based accelerators, demonstrating BETA’s potential for Transformer acceleration at the edge.}
\IEEEpeerreviewmaketitle


\section*{Acknowledgment}

This work was supported by the National Key R\&D Program of China under Grant 2022YFB4400604.



%\bibliographystyle{IEEEtran}

% \clearpage
\normalem
\bibliographystyle{IEEEtran}
\bibliography{ref-jyh}

\begin{thebibliography}{10}
\providecommand{\url}[1]{#1}
\csname url@samestyle\endcsname
\providecommand{\newblock}{\relax}
\providecommand{\bibinfo}[2]{#2}
\providecommand{\BIBentrySTDinterwordspacing}{\spaceskip=0pt\relax}
\providecommand{\BIBentryALTinterwordstretchfactor}{4}
\providecommand{\BIBentryALTinterwordspacing}{\spaceskip=\fontdimen2\font plus
\BIBentryALTinterwordstretchfactor\fontdimen3\font minus \fontdimen4\font\relax}
\providecommand{\BIBforeignlanguage}[2]{{%
\expandafter\ifx\csname l@#1\endcsname\relax
\typeout{** WARNING: IEEEtran.bst: No hyphenation pattern has been}%
\typeout{** loaded for the language `#1'. Using the pattern for}%
\typeout{** the default language instead.}%
\else
\language=\csname l@#1\endcsname
\fi
#2}}
\providecommand{\BIBdecl}{\relax}
\BIBdecl

\bibitem{brown2020language}
T.~Brown, B.~Mann, N.~Ryder \emph{et~al.}, ``Language models are few-shot learners,'' \emph{Advances in neural information processing systems (NeurIPS)}, vol.~33, pp. 1877--1901, 2020.

\bibitem{DevlinCLT19}
J.~Devlin, M.~Chang, K.~Lee \emph{et~al.}, ``{BERT:} pre-training of deep bidirectional transformers for language understanding,'' in \emph{Proceedings of the North American Chapter of the Association for Computational Linguistics: Human Language Technologies (NAACL-HLT)}.\hskip 1em plus 0.5em minus 0.4em\relax ACL, 2019, pp. 4171--4186.

\bibitem{zhou2023solving}
A.~Zhou, K.~Wang, Z.~Lu \emph{et~al.}, ``Solving challenging math word problems using {GPT-4} code interpreter with code-based self-verification,'' \emph{arXiv preprint arXiv:2308.07921}, 2023.

\bibitem{chen2022visualgpt}
J.~Chen, H.~Guo, K.~Yi \emph{et~al.}, ``{VisualGPT:} data-efficient adaptation of pretrained language models for image captioning,'' in \emph{Proceedings of the IEEE/CVF Conference on Computer Vision and Pattern Recognition (CVPR)}, 2022, pp. 18\,030--18\,040.

\bibitem{singh2023progprompt}
I.~Singh, V.~Blukis, A.~Mousavian \emph{et~al.}, ``{ProgPrompt:} generating situated robot task plans using large language models,'' in \emph{IEEE International Conference on Robotics and Automation (ICRA)}.\hskip 1em plus 0.5em minus 0.4em\relax IEEE, 2023, pp. 11\,523--11\,530.

\bibitem{vaswani2017attention}
A.~Vaswani, N.~Shazeer, N.~Parmar \emph{et~al.}, ``Attention is all you need,'' \emph{Advances in neural information processing systems (NeurIPS)}, vol.~30, 2017.

\bibitem{Q8BERT}
O.~Zafrir, G.~Boudoukh, P.~Izsak \emph{et~al.}, ``{Q8BERT:} quantized 8bit {BERT},'' in \emph{Fifth Workshop on Energy Efficient Machine Learning and Cognitive Computing-NeurIPS Edition (EMC2-NeurIPS)}, 2019, pp. 36--39.

\bibitem{TernaryBERT}
W.~Zhang, L.~Hou, Y.~Yin \emph{et~al.}, ``{TernaryBERT:} distillation-aware ultra-low bit {BERT},'' in \emph{Proceedings of the 2020 Conference on Empirical Methods in Natural Language Processing (EMNLP)}.\hskip 1em plus 0.5em minus 0.4em\relax ACL, 2020, pp. 509--521.

\bibitem{BEBERT}
J.~Tian, C.~Fang, H.~Wang \emph{et~al.}, ``{BEBERT}: Efficient and robust binary ensemble {BERT},'' in \emph{IEEE International Conference on Acoustics, Speech and Signal Processing (ICASSP)}.\hskip 1em plus 0.5em minus 0.4em\relax IEEE, 2023, pp. 1--5.

\bibitem{BinaryBERT}
H.~Bai, W.~Zhang, L.~Hou \emph{et~al.}, ``{BinaryBERT:} pushing the limit of {BERT} quantization,'' in \emph{Proceedings of the 59th Annual Meeting of the Association for Computational Linguistics and the 11th International Joint Conference on Natural Language Processing (ACL/IJCNLP)}.\hskip 1em plus 0.5em minus 0.4em\relax ACL, 2021, pp. 4334--4348.

\bibitem{BiT}
Z.~Liu, B.~Oguz, A.~Pappu \emph{et~al.}, ``{BiT:} robustly binarized multi-distilled transformer,'' in \emph{Advances in neural information processing systems (NeurIPS)}, vol.~35, 2022, pp. 14\,303--14\,316.

\bibitem{BiBERT}
H.~Qin, Y.~Ding, M.~Zhang \emph{et~al.}, ``{BiBERT:} accurate fully binarized {BERT},'' in \emph{International Conference on Learning Representations (ICLR)}, 2022.

\bibitem{le2023binaryvit}
P.-H.~C. Le and X.~Li, ``{BinaryViT:} pushing binary vision {Transformers} towards convolutional models,'' in \emph{Proceedings of the IEEE/CVF Conference on Computer Vision and Pattern Recognition (CVPR)}, 2023, pp. 4664--4673.

\bibitem{UCViT}
H.~Song, Y.~Wang, M.~Wang \emph{et~al.}, ``Ucvit: Hardware-friendly vision transformer via unified compression,'' in \emph{IEEE International Symposium on Circuits and Systems (ISCAS)}.\hskip 1em plus 0.5em minus 0.4em\relax {IEEE}, 2022, pp. 2022--2026.

\bibitem{Lu}
S.~Lu, M.~Wang, S.~Liang \emph{et~al.}, ``Hardware accelerator for multi-head attention and position-wise feed-forward in the {Transformer},'' in \emph{IEEE 33rd International System-on-Chip Conference (SOCC)}.\hskip 1em plus 0.5em minus 0.4em\relax {IEEE}, 2020, pp. 84--89.

\bibitem{ViA}
T.~Wang, L.~Gong, C.~Wang \emph{et~al.}, ``{ViA:} {A} novel vision-{Transformer} accelerator based on {FPGA},'' \emph{{IEEE} Trans. Comput. Aided Des. Integr. Circuits Syst. (TCAD)}, vol.~41, no.~11, pp. 4088--4099, 2022.

\bibitem{SwiftTron}
A.~Marchisio, D.~Dura, M.~Capra \emph{et~al.}, ``{SwiftTron:} an efficient hardware accelerator for quantized {Transformers},'' in \emph{International Joint Conference on Neural Networks (IJCNN)}.\hskip 1em plus 0.5em minus 0.4em\relax {IEEE}, 2023, pp. 1--9.

\bibitem{STA}
C.~Fang, A.~Zhou, and Z.~Wang, ``An algorithm-hardware co-optimized framework for accelerating {N:M} sparse {Transformers},'' \emph{{IEEE} Trans. Very Large Scale Integr. Syst. (TVLSI)}, vol.~30, no.~11, pp. 1573--1586, 2022.

\bibitem{EFA-Trans}
X.~Yang and T.~Su, ``{EFA-Trans:} an efficient and flexible acceleration architecture for {Transformers},'' \emph{Electronics}, vol.~11, no.~21, 2022.

\bibitem{FTRANS}
B.~Li, S.~Pandey, H.~Fang \emph{et~al.}, ``{FTRANS:} energy-efficient acceleration of {Transformers} using {FPGA},'' in \emph{Proceedings of the ACM/IEEE International Symposium on Low Power Electronics and Design (ISLPED)}.\hskip 1em plus 0.5em minus 0.4em\relax {ACM}, 2020, pp. 175--180.

\bibitem{ViTA}
S.~Nag, G.~Datta, S.~Kundu \emph{et~al.}, ``{ViTA:} {A} vision {Transformer} inference accelerator for edge applications,'' in \emph{IEEE International Symposium on Circuits and Systems (ISCAS)}.\hskip 1em plus 0.5em minus 0.4em\relax {IEEE}, 2023, pp. 1--5.

\bibitem{GOBO}
A.~H. Zadeh, I.~Edo, O.~M. Awad \emph{et~al.}, ``{GOBO}: Quantizing attention-based {NLP} models for low latency and energy efficient inference,'' in \emph{53rd Annual IEEE/ACM International Symposium on Microarchitecture (MICRO)}.\hskip 1em plus 0.5em minus 0.4em\relax {IEEE}, 2020, pp. 811--824.

\bibitem{A3}
T.~J. Ham, S.~Jung, S.~Kim \emph{et~al.}, ``A\({}^{\mbox{3}}\): Accelerating attention mechanisms in neural networks with approximation,'' in \emph{IEEE International Symposium on High Performance Computer Architecture (HPCA)}.\hskip 1em plus 0.5em minus 0.4em\relax {IEEE}, 2020, pp. 328--341.

\bibitem{fang2022efficient}
C.~Fang, S.~Guo, W.~Wu \emph{et~al.}, ``An efficient hardware accelerator for sparse transformer neural networks,'' in \emph{IEEE International Symposium on Circuits and Systems (ISCAS)}.\hskip 1em plus 0.5em minus 0.4em\relax IEEE, 2022, pp. 2670--2674.

\bibitem{Rodrigue2022Resource-Saving}
R.~Rizk, D.~Rizk, F.~Rizk \emph{et~al.}, ``A resource-saving energy-efficient reconfigurable hardware accelerator for bert-based deep neural network language models using {FFT} multiplication,'' in \emph{IEEE International Symposium on Circuits and Systems (ISCAS)}.\hskip 1em plus 0.5em minus 0.4em\relax {IEEE}, 2022, pp. 1675--1679.

\bibitem{VAQF}
M.~Sun, H.~Ma, G.~Kang \emph{et~al.}, ``{VAQF:} fully automatic software-hardware co-design framework for low-bit vision {Transformer},'' \emph{arXiv preprint arXiv:2201.06618}, 2022.

\bibitem{FQ-BERT}
Z.~Liu, G.~Li, and J.~Cheng, ``Hardware acceleration of fully quantized {BERT} for efficient natural language processing,'' in \emph{Design, Automation \& Test in Europe Conference (DATE)}.\hskip 1em plus 0.5em minus 0.4em\relax {IEEE}, 2021, pp. 513--516.

\bibitem{ITA}
G.~Islamoglu, M.~Scherer, G.~Paulin \emph{et~al.}, ``{ITA:} an energy-efficient attention and softmax accelerator for quantized {Transformers},'' in \emph{Proceedings of the ACM/IEEE International Symposium on Low Power Electronics and Design (ISLPED)}.\hskip 1em plus 0.5em minus 0.4em\relax {IEEE}, 2023, pp. 1--6.

\bibitem{DBLP:journals/corr/abs-2306-06446}
H.~You, H.~Shi, Y.~Guo \emph{et~al.}, ``{ShiftAddViT:} mixture of multiplication primitives towards efficient vision {Transformer},'' in \emph{Advances in neural information processing systems (NeurIPS)}, vol.~36, 2023.

\bibitem{GLUE}
A.~Wang, A.~Singh, J.~Michael \emph{et~al.}, ``{GLUE:} {A} multi-task benchmark and analysis platform for natural language understanding,'' in \emph{International Conference on Learning Representations (ICLR)}, 2019.

\end{thebibliography}

\end{document}